\documentclass{article}
\usepackage{amsmath}

\begin{document}

\title{Influence of Gravity on noncommutative Dirac equation}
\author{S. Bourouaine \thanks{
phybou@wissal.dz} \, {\text{and}} A. Benslama \\
Physics Department, Faculty of Sciences, Mentouri University, \\
Constantine, Algeria}
\maketitle

\begin{abstract}
In this paper, we investigate the influence of gravity and noncommutativity
on Dirac equation. By adopting the tetrad formalism, we show that the
modified Dirac equation keeps the same form. The only modification is in the
expression of the covariant derivative. The new form of this derivative is
the product of its counterpart given in curved space-time with an operator
which depends on the noncommutative $\theta$-parameter. As an application,
we have computed the density number of the created particles in presence of
constant strong electric field in an anisotropic Bianchi universe.
\end{abstract}

\markboth{S. Bourouaine and A. Benslama} {(Influence of Gravity on
noncommutative Dirac equation)}

\section{Introduction}

Recently there has been a large interest in the study of noncommutative
field theory. Taking space-time coordinates to be noncommutative is an old
idea dates back to the work of Snyder\cite{[1]}. The goal was that the
introduction of a noncommutative structure to space-time at small length
scales could introduce an effective cut-off which regularize divergences in
quantum field theory. However this theory was plagued with several problems
such as the violation of unitarity, causality, etc.. which make people
abandon it. Recently, the appearance of such theories as certain limits of
string theory, D-brane and M-theory have generated a revival of interest for
field theory on a noncommutative space-time\cite{[2],[3]}. For a review of
noncommutative field theories see\cite{[4]}. \newline
The noncommutative space-time is characterized by operators $\hat{x}_{\mu }$
satisfying the relation
\begin{equation}
\lbrack \hat{x}^{\mu },\hat{x}^{\nu }]=i\theta ^{\mu \nu }=\frac{i}{\Lambda
_{NC}^{2}}C^{\mu \nu }
\end{equation}%
where $\theta ^{\mu \nu }$ is antisymmetric matrix which has the dimension
of area. Its role can be compared to that of the Planck constant $h$ which
quantifies in quantum mechanics the level of noncommutativity between space
and momentum. $C^{\mu \nu }$are dimensionless parameters which are
presumably of order unity, and $\Lambda _{NC}$ is the scale energy where the
noncommutative effects of the space-time become relevant. At the beginning
this scale was expected to manifest around the Planck scale which makes the
noncommutative effects of space-time out of reach of the current colliders.
However if we assume the possibility of large extra dimensions, it is likely
that $\Lambda _{NC}$ could be set at a TEV scale\cite{[5]}. \newline
There are two different settings for noncommutative field theories. The
first one is based on the so-called Moyal or star product. The other setting
uses the Sieberg-Witten map\cite{[3],[4],[5],[6]}. In this work we shall
adopt the first approach where field theory in a noncommutative space-time
can be described as the ordinary field theory where every product of fields
is replaced with the Moyal or star product $(\ast )$ defined as
\begin{equation}
f(x)\ast g(x)=[\exp (\frac{i}{2}\theta ^{\mu \nu }\partial _{\mu }^{(\xi
)}\partial _{\nu }^{(\eta )})f(x+\xi )g(x+\eta )]_{\xi =\eta =0}
\end{equation}%
where $f$ and $g$ are two arbitrary differentiable fields. This deviation
from the standard theory causes the violation of Lorentz invariance when $%
\theta $ is considered as a constant matrix, except if this matrix is
promoted to a tensor related to the contracted Snyder's Lie algebra\cite%
{[7],[8]}. Also, it has been shown that the problem of unitarity appears in
the study of quantum theories in flat space with time-space
noncommutatitvities $(\theta ^{0i}\neq 0)$\cite{[9],[10]}. Most applications
of ordinary field theories have been reconsidered in the context of
noncommutative geometry \cite{[11],[12],[13],[14]}. Among these
applications, the pair production in the presence of electric field has been
treated by a nonperturbative theory with a noncommutative Dirac equation in
the absence of gravity\cite{[15]}. The second section is devoted to
establishing the modified Dirac equation in curved space-time. This
modification is due to the influence of noncommutativity and gravity on
Dirac particle. In fact, we will adopt the tetrad formalism\cite{[16]} as a
tool to formulate the final equation up to the first order in the
noncommutative parameter $\theta $. As an application, we will propose in
the third section an example of pair production in the presence of strong
electric field in the Bianchi 1 universe.

\section{Modified Dirac Equation in curved space}

It is well known that to determine the effects of gravitation on
general dynamical system, one has just to take the
special-relativistic motion equations in the absence of
gravitation and replace all components of the Lorentz tensors with
its components given in non flat space. Also, we replace all
derivatives $\partial _{\alpha }$ with covariant derivatives and
the Minkowski metric $\eta _{\alpha \beta }$ with $g_{\alpha \beta
}$. It is possible to determine the effects of gravity on spinor
field by using the tetrad formalism\cite{[16]} which is based on
the principle of equivalence. The metric in any general non
inertial coordinate system is given by
\begin{equation}
g_{\mu \nu }(x)=e_{\mu }^{(\alpha )}e_{\nu }^{(\beta )}(x)\eta _{\alpha
\beta }\quad and\quad e_{\mu }^{(\alpha )}=\partial _{\mu }\zeta
_{x}^{(\alpha )}
\end{equation}%
$e_{{\mu }}^{(\alpha )}(x)$ are the vierbeins or tetrads that connect
between the curved space and its local flat counterpart. Its inverse is $%
e_{(\alpha )}^{{\mu }}=\frac{\partial x^{\mu }}{\partial \xi _{x}^{\alpha }}$
such that
\begin{equation}
e_{(\alpha )}^{{\mu }}e_{\nu }^{(\alpha )}=\delta _{\nu }^{\mu }
\end{equation}%
We can use the vierbeins to refer the components of contravariant vector $T^{%
{\mu }}(x)$ into the local coordinate system $\xi _{x}^{\alpha }$
\begin{equation}
T^{\mu }\rightarrow e_{\alpha }^{(\mu )}T^{\alpha }
\end{equation}%
Generally, for any tensor $T_{\beta }^{\alpha }$ we have
\begin{equation}
T_{\beta }^{\alpha }\rightarrow e_{\mu }^{(\alpha )}e_{(\beta )}^{\nu
}T_{\nu }^{\mu }
\end{equation}%
Now, we consider that the flat space time is deformed by $\theta $
-parameter which makes their coordinates do not commute given by (1) and we
take $\theta ^{\alpha \beta }=\theta \epsilon ^{\alpha \beta }$$(\epsilon
^{\alpha \beta }$ is Levi-Cevita tensor). \newline
To study the influence of gravity on Dirac equation given in a deformed
space time (1), we start in the first step by giving the noncommutative
Dirac equation that describes spinorial particle in the presence of
electromagnetic field and the absence of gravity\cite{[15]}
\begin{equation}
\lbrack \gamma ^{\mu }(\partial _{\mu }-ieA_{\mu }(\hat{x}))+m]\psi (\hat{x}%
)=0,\quad with\quad \lbrack \hat{x}^{\mu },\hat{x}^{\nu }]=i\theta ^{\mu \nu
},
\end{equation}%
where $\gamma ^{\mu }$ are the usual Dirac matrices that satisfy $[\gamma ^{{%
\mu }},\gamma ^{\nu }]=2\eta ^{\mu \nu }$. Choosing
\begin{align}
& \gamma ^{0}=\left(
\begin{array}{cc}
-i\sigma _{1} & 0 \\
0 & i\sigma _{2}%
\end{array}%
\right) ,\gamma ^{1}=\left(
\begin{array}{cc}
0 & i \\
-i & 0%
\end{array}%
\right) ,  \notag \\
& \gamma ^{2}=\left(
\begin{array}{cc}
\sigma _{2} & 0 \\
0 & \sigma _{2}%
\end{array}%
\right) ,\gamma ^{3}=\left(
\begin{array}{cc}
\sigma _{3} & 0 \\
0 & -\sigma _{3}%
\end{array}%
\right)
\end{align}%
$\sigma _{1},\sigma _{2}$ and $\sigma _{3}$ are Pauli matrices.\newline
By using the multiplication law of star product (2) up to the first order
into eq. (7), we get
\begin{equation}
\lbrack \gamma ^{\mu }(\partial _{\mu }-ieA_{\mu }+\frac{e}{2}\theta
^{\alpha \rho }(\partial _{\alpha }A_{\mu })\partial _{\rho })+m]\psi =0
\end{equation}%
due to the smallness of the $\theta $-parameter $(\theta \ll 1)$. As a
consequence, the eq. (7) has an analogical equation in the commutative space
\begin{equation}
\lbrack \gamma ^{\mu }(D_{\mu }(x)-ieA_{\mu }(x))+m]\psi =0,\quad with\quad
\lbrack x^{\mu },x^{\nu }]=0,
\end{equation}%
where $D_{\mu }$ is the modified derivative
\begin{equation}
D_{\mu }=M_{\mu }^{\rho }(\theta )\partial _{\rho }\quad with\quad M_{\mu
}^{\rho }(\theta )=\delta _{\mu }^{\rho }+\frac{e}{2}\theta ^{\alpha \rho
}(\partial _{\alpha }A_{\mu }).
\end{equation}%
The part which does not contain the $\theta $-dependent corrections and
corresponds to the usual Dirac operator remains covariant. \newline
Then, similar to the commutative case, and in order to establish the
modified Dirac equation in the presence of gravity, we follow the tetrad
formalism approach and make the following changes in eq.(10)\cite{[16]}
\begin{align}
& A_{\mu }\rightarrow e_{(\mu )}^{\alpha }A_{\alpha },  \notag \\
& \partial _{\alpha }A_{\mu }\rightarrow e_{(\alpha )}^{\nu }e_{(\mu
)}^{\alpha }(\partial _{\nu }A_{\alpha }),  \notag \\
& \partial _{\mu }\psi \rightarrow e_{(\mu )}^{\alpha }(\partial _{\alpha
}-\Gamma _{\alpha })\psi ,
\end{align}%
where $\Gamma _{\mu }$ is the spin connection, its role is to conserve the
covariance that exist already in the commutative part (does not depend on
matrix $\theta $) in order not to make a conflict in the classical limit
when ($\theta \rightarrow 0$)
\begin{equation}
\Gamma _{\mu }=\frac{1}{4}g_{\lambda \alpha }[(\partial _{\mu }e_{\nu
}^{(\beta )})e_{(\beta )}^{\alpha }-\Gamma _{\nu \mu }^{\alpha }]s^{\lambda
\nu }
\end{equation}%
with
\begin{equation}
s^{\lambda \nu }=\frac{1}{2}e_{(\rho )}^{\mu }e_{(\sigma )}^{\nu }[\gamma
^{\rho }\gamma ^{\sigma }-\gamma ^{\sigma }\gamma ^{\rho }]
\end{equation}%
and $\Gamma _{\nu \mu }^{\alpha }$ is the affine connection, which is
written in function of the metric $g_{\lambda \alpha }$ as
\begin{equation}
\Gamma _{\nu \mu }^{\alpha }(x)=(1/2)g^{\alpha \beta }(\partial _{\nu
}g_{\mu \beta }+\partial _{\mu }g_{\beta \nu }-\partial _{\beta }g_{\mu \nu
}).
\end{equation}%
Finally, by substituting the changes given in eqs. (12) into the modified
derivative (11), we get
\begin{equation}
D_{\mu }\rightarrow e_{(\mu )}^{\alpha }\hat{\mathit{D}}_{\alpha }
\end{equation}%
where $\hat{\mathit{D}}_{\alpha }$ is the modified Dirac derivative in the
presence of gravity
\begin{equation}
\hat{\mathit{D}}_{\mu }=\hat{M}_{\mu }^{\rho }(\theta )(\partial _{\rho
}-\Gamma _{\rho }),
\end{equation}%
with
\begin{equation}
\hat{M}_{\mu }^{\rho }(\theta )=\delta _{\mu }^{\rho }+\frac{e}{2}\theta
^{\alpha \rho }e_{(\alpha )}^{a}e_{(\mu )}^{b}(\partial _{a}A_{b})
\end{equation}%
the modified Dirac equation in the presence of gravity is
\begin{equation}
\lbrack \tilde{\gamma}^{\alpha }(\hat{\mathit{D}}_{\alpha }-ieA_{\alpha
})+m]\psi =0.
\end{equation}%
The matrices $e_{(\nu )}^{\mu },e_{\mu }^{(\nu )}$ connect between the $%
\tilde{\gamma}^{\mu }$ and Minkowski $\gamma ^{\mu }$ Dirac matrices as
follow:
\begin{equation}
\tilde{\gamma}^{\alpha }=\gamma ^{\mu }e_{(\mu )}^{\alpha }
\end{equation}%
$\tilde{\gamma}^{\alpha }$ satisfy the anticommutation relation
\begin{equation}
\lbrack \tilde{\gamma}^{\alpha },\tilde{\gamma}^{\beta }]_{+}=2g^{\alpha
\beta }.
\end{equation}

\section{Creation of Dirac particles in the presence of a constant electric
field in an anisotropic Bianchi 1 universe}

The creation of particles induced by the vacuum instability is among the
interesting non perturbative phenomena which can be realized with a presence
of strong electromagnetic field. \newline
In this section, we will see an example of Dirac particles created in a
cosmological anisotropic Bianchi 1 universe in the presence of a strong
constant electric field using the modified Dirac equation (19), in order to
deduce the effect of the noncommutativity on the density of created
particles. \newline
The line element that defines the Bianchi universe is given by
\begin{equation}
ds^{2}=-dt^{2}+t^{2}(dx^{2}+dy^{2})+dz^{2},
\end{equation}%
then the metric
\begin{equation}
g_{\mu \nu }=\left(
\begin{array}{cccc}
-1 & 0 & 0 & 0 \\
0 & t^{2} & 0 & 0 \\
0 & 0 & t^{2} & 0 \\
0 & 0 & 0 & 1%
\end{array}%
\right)
\end{equation}%
and its inverse is
\begin{equation}
g^{\mu \nu }=\left(
\begin{array}{cccc}
-1 & 0 & 0 & 0 \\
0 & \frac{1}{t^{2}} & 0 & 0 \\
0 & 0 & \frac{1}{t^{2}} & 0 \\
0 & 0 & 0 & 1%
\end{array}%
\right) .
\end{equation}%
Since the line element is diagonal, we choose to work in the diagonal tetrad
\begin{equation}
e_{(\alpha )}^{\mu }=\sqrt{|g^{\mu \beta }|\delta _{\alpha \beta }},
\end{equation}%
from (13), (22) and (24), we get all components of spin connection
\begin{equation}
\Gamma _{1}=\frac{1}{2}\gamma ^{0}\gamma ^{1},\Gamma _{2}=\frac{1}{2}\gamma
^{0}\gamma ^{2},\Gamma _{3}=\Gamma _{0}=0
\end{equation}%
We fix the third axis of our frame parallel to the direction of constant
electric field $\overrightarrow{E}$. This means that the system possesses a
rotational symmetry along $z$ axis that preserves the invariance of line
element given by (22). In such cases, the electric field has only one
component $E_{z}=E$ which corresponds to the chosen potential
\begin{equation}
A_{\mu }=(0,0,0,-Et).
\end{equation}%
The modified Dirac equation (19) in anisotropic Bianchi universe (22) with
the presence of electric field represented by the potential $A_{\mu }$ in
(27) is
\begin{equation}
\lbrack \gamma ^{\alpha }e_{(\alpha )}^{\mu }((\delta _{\mu }^{\rho }+\frac{e%
}{2}\theta ^{0\rho }\delta _{\mu }^{3}E)(\partial _{\rho }-\Gamma _{\rho
})-ieA_{\mu })+m]\psi =0
\end{equation}%
To visualize the influence of the noncommutativity on the creation of Dirac
particles, we use the parametrization that determines the elements of $%
\theta $-Matrix from the direction of background electromagnetic fields\cite%
{[11]}
\begin{equation*}
C^{\mu \nu }=\left(
\begin{array}{cccc}
0 & \sin \alpha \cos \beta  & \sin \alpha \sin \beta  & \cos \alpha  \\
-\sin \alpha \cos \beta  & 0 & \cos \gamma  & -\sin \gamma \sin \beta  \\
-\sin \alpha \sin \beta  & -\cos \gamma  & 0 & -\sin \gamma \cos \beta  \\
-\cos \alpha  & \sin \gamma \sin \beta  & \sin \gamma \cos \beta  & 0%
\end{array}%
\right)
\end{equation*}%
and again keep the background electric field parallel to $z$ axis to benefit
from the existed rotational symmetry in order to be sure that the density
number of particles remains unchanged under this symmetry. Although, the
Lorentz symmetry is generally broken when $\theta $-matrix is not considered
as a tensor and all its elements are identical in all reference frames.
Geometrically, we have $\alpha =0$ to obtain $\theta ^{0i}=\theta
^{03}=\theta $, and set all the remaining time space components equal to
zero), then the eq (28) becomes
\begin{equation}
\lbrack \gamma ^{0}\frac{\partial }{\partial t}+\gamma ^{3}(e^{\frac{e}{2}%
\theta E}\frac{\partial }{\partial z}+ieEt)+\frac{1}{t}(\gamma ^{1}\frac{%
\partial }{\partial x}+\gamma ^{2}\frac{\partial }{\partial y})+m]\widehat{%
\psi }=0.
\end{equation}%
We have used $\widehat{\psi }=t\psi $ to cancel the term related to the spin
connection in the equation (28), and have taken
\begin{equation}
e^{\frac{e}{2}\theta E}\approx (1+\frac{e}{2}\theta E).
\end{equation}%
Since the operators $(\partial _{x},\partial _{y},\partial _{z})$ commute
with the operator acting on $\widehat{\psi }$, the general solution is
\begin{equation}
\widehat{\psi }=\exp (\overrightarrow{k}.\overrightarrow{r})\left(
\begin{array}{c}
\varphi (t) \\
\chi (t)%
\end{array}%
\right)
\end{equation}%
where $\varphi (t),\chi (t)$ are the vectors which depend only on time.
\begin{equation}
\overrightarrow{r}=(x,y,z)\quad and\quad \overrightarrow{k}%
=(ik_{x},ik_{y},ik_{z}).
\end{equation}%
The operator acting on $\Psi =\gamma ^{3}\gamma ^{0}\widehat{\psi }$ can be
written as the sum of two commuting operators $\hat{O}_{1}$ and $\hat{O}_{2}$
\begin{equation}
\lbrack \hat{O}_{1},\hat{O}_{2}]=0,
\end{equation}%
such that
\begin{equation}
\frac{\hat{O}_{1}}{t}=[\gamma ^{3}\frac{\partial }{\partial t}+i\gamma
^{0}(e^{\frac{e}{2}\theta E}k_{z}+eEt)+\gamma ^{3}\gamma ^{0}m]
\end{equation}%
\begin{equation}
\hat{O}_{1}=i(\gamma ^{1}k_{x}+\gamma ^{2}k_{y})\gamma ^{3}\gamma ^{0}
\end{equation}%
By considering $\Psi $ as an eigenvector of $\hat{O}_{2}$ with eigenvalues $k
$, we get
\begin{equation}
\hat{O}_{2}=\left(
\begin{array}{c}
\varphi (t) \\
\chi (t)%
\end{array}%
\right)
\begin{array}{c}
\end{array}%
=k\left(
\begin{array}{c}
\varphi (t) \\
\frac{k_{x}\sigma _{2}}{ik_{y}-k}\varphi (t)%
\end{array}%
\right)
\begin{array}{c}
\end{array}%
\end{equation}%
\begin{equation}
k=i\sqrt{k_{x}^{2}+k_{y}^{2}}=ik_{\bot },
\end{equation}%
and
\begin{equation}
\varphi (t)=\left(
\begin{array}{c}
f_{1}(t) \\
f_{2}(t)%
\end{array}%
\right)
\begin{array}{c}
\end{array}%
.
\end{equation}%
On the other hand, we have $\Psi $ is an eigenvector of $\hat{O}_{1}$, then $%
f_{1}(t)$ and $f_{2}(t)$ satisfy the following differential equations system
\begin{equation}
\frac{\partial f_{1}(t)}{\partial t}+\frac{k}{t}f_{1}(t)+(eEt+\widetilde{k}%
_{z}+im)f_{2}(t)=0
\end{equation}%
\begin{equation}
-\frac{\partial f_{2}(t)}{\partial t}+\frac{k}{t}f_{2}(t)+(eEt+\widetilde{k}%
_{z}-im)f_{1}(t)=0
\end{equation}%
where the modified vector wave
\begin{equation}
\widetilde{k}_{z}=k_{z}e^{\frac{e}{2}\theta E}
\end{equation}%
Physically, the created particles undergo an added acceleration to their
impulsions along the direction of background electric field. \newline
In order to determine the density number of created particles, we should
take into account the asymptotic behavior of $f_{1}$ and $f_{2}$ at $%
t\rightarrow 0$ and $t\rightarrow \infty $. \newline
By considering the ultra relativistic approximation $|E|\gg m,k_{z}$ and
assuming that $|\theta |\ll |\frac{m}{E}|$, while neglecting $|\frac{k_{z}}{E%
}|$ and $|\frac{m}{E}|$ in the first order variation of $f_{1}(t)$ and $%
f_{2}(t)$; then eqs. (37), (38) reduce to the same solutions form as that
given without the influence of the noncommutativity\cite{[17]}, i.e. we
obtain the solutions in terms of Whittaker functions
\begin{equation}
\frac{f_{2}(t)}{\sqrt{t}}=[C_{1}M_{\lambda ^{+},\mu
}(ieEt^{2})+C_{2}W_{\lambda ^{+},\mu }(ieEt^{2})]
\end{equation}%
\begin{equation}
\frac{f_{1}(t)}{\sqrt{t}}=[C_{3}M_{\lambda ^{-},\mu
+1}(ieEt^{2})+C_{4}W_{\lambda ^{-},\mu +1}(ieEt^{2})]
\end{equation}%
\begin{equation}
\lambda ^{\pm }=\frac{i(im\pm \widetilde{k}_{z})^{2}}{4eE},\mu =\frac{k}{2}-%
\frac{1}{2},
\end{equation}%
where C1,C2 and C3,C4 are arbitrary constants. \newline
We compute the density number of particles creation with the help of the
Bogoliubov coefficients $\alpha ,\beta $\cite{[17],[18],[19]} which relate
between the limit of the negative frequency solutions $\psi _{0}^{-}$ at $%
t\rightarrow 0$ and both $\psi _{\infty }^{-}$(the limit negative frequency
solution at $t\rightarrow \infty $) and $\psi _{\infty }^{+}$( the limit of
the positive frequency solution at $t\rightarrow \infty $), knowing that the
negative or positive frequency mode is extracted by using the sign of the
operator $i\partial _{t}$
\begin{equation}
\psi _{0}^{-}=\alpha \psi _{\infty }^{-}+\beta \psi _{\infty }^{+}.
\end{equation}%
By following the same steps given in\cite{[17]}, we deduce the result $(%
\frac{|\beta |^{2}}{|\alpha |^{2}})_{NC}$
\begin{equation}
(\frac{|\beta |^{2}}{|\alpha |^{2}})_{NC}=e^{2i\pi \mu }\frac{|\Gamma (\frac{%
1}{2}+\mu -\lambda ^{\pm })|^{2}}{|\Gamma (\frac{1}{2}+\mu +\lambda ^{\pm
})|^{2}} \\
=e^{-\pi k_{\bot }}\frac{(\frac{k_{\bot }}{2}+\tilde{\lambda})\sinh \pi (%
\frac{k_{\bot }}{2}+\tilde{\lambda})}{(\frac{k_{\bot }}{2}-\tilde{\lambda}%
)\sinh \pi (\frac{k_{\bot }}{2}-\tilde{\lambda})}
\end{equation}%
where we have considered that
\begin{equation}
(\frac{k_{\bot }}{2}+\tilde{\lambda})=Im(\frac{1}{2}+\mu +\lambda ^{\pm
})\gg Re(\frac{1}{2}+\mu +\lambda ^{\pm }),
\end{equation}%
and
\begin{equation}
\tilde{\lambda}=-\frac{m^{2}-\widetilde{k}_{z}^{2}}{4eE}.
\end{equation}%
The density number of created particles with the influence of
noncommutativity is
\begin{equation}
n_{NC}=[(\frac{|\beta |^{2}}{|\alpha |^{2}})_{NC}^{-1}+1].
\end{equation}%
Remark that the final result of density number of particles creation (49) is
independent from the rotational symmetry of space along $z$ axis. This means
that our frame has only the particular symmetries and is fixed by the
background field. This point of view is coming from the first approach of
noncommutative space time\cite{[3],[4],[5],[6]} based on constant $\theta $%
-matrix. Finally by using the perturbative expansion of (49) up to the first
order of $\theta $-parameter, we get
\begin{equation}
n_{NC}=n+\Delta n
\end{equation}%
$n$denotes the usual density number of created particles. \newline
$\Delta n$ denotes the correction due to the noncommutativity such that
\begin{equation}
\Delta n=\frac{\theta k_{z}^{2}}{4}e^{-\pi k_{\bot }}[\frac{\sinh (\pi
k_{\bot })}{\sinh ^{2}(\frac{\pi k_{\bot }}{2}+\frac{\pi m^{2}}{4eE})}+\frac{%
k_{\bot }}{(\frac{k_{\bot }}{2}+\frac{m^{2}}{4eE})^{2}}].
\end{equation}%
It is clear that when $\theta \rightarrow 0$, we have $\Delta n\rightarrow 0$%
. \newline
In the absence of the electric field $(E\rightarrow 0)$ we can show easily
that $\Delta n\rightarrow 0$ and we recover that the density of created
particle becomes thermal\cite{[20]}
\begin{equation}
n_{NC}\approx n\approx e^{-2\pi k_{\bot }}.
\end{equation}

\section{Conclusion}

We have deduced the form of the Dirac equation in the presence of gravity in
the framework of a noncommutative space-time. \newline
By adopting the tetrad formalism, we have shown that the modified Dirac
equation keeps the standard form except a modification in the expression of
the covariant derivative. The new form of this derivative is the product of
its counterpart in curved space-time by an operator which depends on the
noncommutative $\theta$ parameter. As an application, we have computed the
density number of the created particles in presence of constant strong
electric field in an anisotropic Bianchi universe. Our calculation shows a
correction due to the noncommutativity to be contrasted with the work of
Chair et. al. where it was found that there is no correction induced by
noncommutativity on a flat space-time. The main perspective of this work is
to use $\theta$-matrix as a tensor to preserve the Lorentz covariance in the
same spirit of reference\cite{[7]}.

\end{document}